# Molecular dynamics simulation of UO$_2$ nanocrystals surface

A.S. Boyarchenkov[a], S.I. Potashnikov[a], K.A. Nekrasov[a], A.Ya. Kupryazhkin[a]

[a] Ural Federal University, 620002, Mira street 19, Yekaterinburg, Russia

boyarchenkov@gmail.com  potashnikov@gmail.com  kirillnkr@mail.ru  kupr@dpt.ustu.ru

**Abstract**

In this article we investigated surface of nanocrystals (NC) of uranium dioxide (UO$_2$) using molecular dynamics (MD) under isolated (non-periodic) boundary conditions with the approximation of pair potentials and rigid ions. It is shown that a cubic shape of the model NCs is metastable and the stable equilibrium is reached in the process of structural relaxation to the octahedral shape over a time of 1000 ns (200 million MD steps), which increases with the size of NC. We measured the size dependences of the lattice parameter and the surface energy density for NC of cubic and octahedral shape with volume up to 1000 nm$^3$ (50000 particles) at temperatures of 2200K and 2300K. For the surfaces {100} and {111} we obtained the energy density $\sigma_{100}=1.602\pm0.016$ J/m$^2$, $\sigma_{111}=1.137\pm0.032$ J/m$^2$ and surface tension constant $\gamma_{111}=0.875\pm0.008$ J/m$^2$. The resulting ratio of $\sigma_{100}/\sigma_{111}=1.408\pm0.042$ within the error coincides with the experimental value of $1.42\pm0.05$ measured for microscopic cavities in monocrystals of UO$_2$.

## 1. Introduction

The surface structure of uranium dioxide (UO$_2$) is important in the manufacture of nuclear fuel due to the influence of surface energy on the morphology of the powder (in equilibrium the surface of the lower energy is more common). Similarly, the segregation of fission products can be suppressed or accelerated depending on the configuration of surface ions.

For experimental study of surface characteristics of UO$_2$ the spectroscopic and microscopic techniques are used (see review in [1]). Quantum-mechanical (*ab initio*) calculations allow one to track changes in the electronic structure depending on the distance from the surface, but due to large computational complexity such calculations are limited to a small number of ~100 atoms [2] [3], even when the approximations of the density functional theory (DFT) are used. On the other hand, empirical interaction potentials have manifested themselves well in molecular dynamics (MD) simulation of quasi-infinite periodic crystals (see the reviews [4] [5] [6]), but their adequacy for modeling the surface properties must be examined specially.

Computer simulation of UO$_2$ surfaces had previously been performed both on the basis of pair potentials (SPP-calculations) in the shell model [2] [7] and the density functional theory (DFT-calculations) [2] [3]. However, these calculations used quasi-infinite crystals with 2D or 3D-periodicity along with the method of lattice statics (LS) without particle dynamics and, therefore, without consideration of kinetic effects.

In this paper, we for the first time used MD simulation of non-periodic nanocrystals (NC) of cubic and octahedral shapes (CNC and ONC) up to 1000 nm$^3$ (50000 particles) for investigation of the surface properties of UO$_2$ by analyzing the dependencies of the lattice constant and energy density on the reciprocal linear size. Simulation time required to obtain the equilibrium values of macroscopic parameters of the nanocrystals (10–100 thousand particles) under isolated boundary conditions (IBC) is significantly higher than the corresponding times for quasi-infinite crystals (about 1000 particles) under periodic boundary conditions (PBC). Therefore, we restricted ourselves to simulation with the only one set of pair potentials (SPP) MOX-07 [8], which reproduces the largest specter of known experimental data for UO$_2$ [6].

## 2. Methodology

The original model of this work is non-periodic nanocrystal of uranium dioxide, surrounded by vacuum (i.e. under IBC), which is formed of $N$ rigid ions using the face centered cubic (FCC) unit cell with zero dipole moment. The interaction between particles was described by pair potentials with the Coulomb term and the short-range Buckingham term [9]. The resulting forces and system energy were calculated as superposition of $N(N-1)/2$ independent pair interactions:

$$\mathbf{r}_{ij} = \mathbf{r}_i - \mathbf{r}_j; \quad r_{ij} = \sqrt{\mathbf{r}_{ij}\mathbf{r}_{ij}}$$

$$U_{ij} = \frac{Q^2 K_e q_i q_j}{r_{ij}} + X_{ij}\exp(-Y_{ij}r_{ij}) - \frac{Z_{ij}}{r_{ij}^6}$$

$$\mathbf{f}_i = -\sum_{j=1}^{N}\nabla U_{ij}; \quad E = \sum_{i=1}^{N}\frac{m_i \mathbf{v}_i^2}{2} + \sum_{i=1}^{N}\sum_{j\neq i}^{N}\frac{U_{ij}}{2}.$$

Here $\mathbf{r}_i$, $\mathbf{f}_i$, $\mathbf{v}_i$, $q_i$, $m_i$ are the position, force, velocity, charge and mass of the $i$-th particle; $U_{ij}$ is the pair potential with dimensionless ionicity coefficient $Q$ and the parameters of short-range $X$, $Y$, $Z$, which are shown in Table 1 (X++, Y++, Z++ and Z+- equal zero); $K_e$ = 14.399644 eV*Å is the electrostatic constant; $E$ is the total energy of the system.

In order to integrate the Newton equations of motion we used the reversible semi-implicit Euler method [10] with a time step of 5 fs, correction of displacement of the center of mass and rotation around it,

as well as quasicanonical dissipative Berendsen thermostat [11] with a relaxation time of 10 ps:

$$x(t) = \sqrt{1 + \left(\frac{T_{stat}}{T_{system}(t)} - 1\right) \cdot \frac{1}{\tau_T}}$$

$$\mathbf{v}'_i(t+\Delta t) = \left(\mathbf{v}_i(t) + \frac{\mathbf{f}_i(t)}{m_i} \cdot \Delta t\right) - \frac{\sum_{j=1}^{N} m_j \cdot \mathbf{v}_j(t)}{\sum_{j=1}^{N} m_j}$$

$$\mathbf{v}_i(t+\Delta t) = \left(\mathbf{v}'_i(t+\Delta t) - \mathbf{W}(\mathbf{r}_i(t), \mathbf{v}'_i(t+\Delta t)) \times \mathbf{r}_i(t)\right) \cdot x(t)$$

$$\mathbf{r}_i(t+\Delta t) = \mathbf{r}_i(t) + \mathbf{v}_i(t+\Delta t) \cdot \Delta t.$$

Here $T_{system}$ is the instantaneous temperature of the system; $T_{stat}$, $\tau_T$, $x$ are the temperature, relaxation time and coefficient of the thermostat; $\mathbf{W}$ is the angular velocity of the system after the correction of displacement of the center of mass.

At each step of the dynamics we calculated the instantaneous numerical density of particles $n(t)$ in the system by averaging the density over an ensemble of $S$ spherical layers. Then, we calculated an average lattice parameter $<a>$ from the time-averaged density and the constant number of particles per FCC unit cell:

$$n(t) = \frac{1}{S}\sum_{s=1}^{S}\frac{N_s(t)}{V_s(t)}; \quad \langle a \rangle = \left(\frac{12}{\langle n(t) \rangle}\right)^{1/3}$$

All the MD-simulations were carried out on graphics processors (GPU) with NVIDIA CUDA technology, which gave us speedup of 2–3 orders (see details in our papers [12] [13]).

### 3. Nanocrystals shape relaxation

Since we form the initial configuration of NC replicating the unit FCC cell, the particles on its surface are unbalanced due to asymmetric charge environment (they lack a few neighbors). After a brief MD simulation (about 0.1 ns), the surface reaches a state of metastable equilibrium, clinging to the interior of the crystal. This process is known as kinetic relaxation. Then, after a prolonged MD simulation (about 1000 ns) the stable equilibrium of the entire crystal is reached. In this case, NC of cubic shape transforms to the octahedral NC via surface diffusion and there is contraction of the inner region due to surface tension forces. This process is called structural relaxation. Fig. 1 and 2, for example, show the evolution of the lattice constant and energy of CNC of 6144 particles in the process of such structural relaxation, which is stepwise, as one can see.

Thus, the CNC of cubic shape confined by crystallographic planes of the type {100}, were energetically less favorable compared to ONC confined by planes {111}. Fig. 3 shows the cationic (uranium) sublattice of NC in the beginning of the simulation and after the relaxation. It is seen that the relaxed crystal differs from the regular octahedron only in smoothed corners. In modeling of crystallization process the grown NC also had a shape of an octahedron.

In principle, the equilibrium shape of single crystals should be determined by the crystallographic planes with a minimum surface energy. Indeed, monocrystals of $UO_2$ often take the form of octahedra (see, for example, [14]). However, the external shape of crystals is often determined not so much by the minimum surface energy, as the growth conditions (interaction of the surface with environment) and the presence of impurities. Therefore, the final conclusion on the optimal shape of nanocrystals can not be based on the exterior form of natural crystals. In this sense, more reliable references are the evidence of cleavage [15] and the equilibrium shape of internal microscopic cavities (bubbles) in single crystals, since they are not influenced by the external environment, and the equilibrium shape is reached fast enough.

It is shown in the experimental work of Castell [16], that such cavities in $UO_2$ single crystals have a shape close to octahedral: their faces are formed by planes {111} and the corners are smoothed by planes {100} (see Fig. 4). Thus, our results of MD simulation, which show the formation of octahedral NC having smooth corners (see Fig. 3), are in a very good agreement with experimental data.

For additional verification of the equilibrium state, we simulated NCs having the octahedral shape from the beginning. We used the following algorithm for creating ONC of any given number of particles:

- Create CNC with a linear size of 2C unit FCC cells on the edge (i.e., a crystal with diameter larger than the diameter of the required octahedron);
- Rearrange the ions randomly (to ensure that removing of redundant ions would not break the symmetry);
- Sort ions in ascending order of the norm $||R|| = |R_x| + |R_y| + |R_z|$ (considering that the equation $||R|| = $ Const defines the surface of octahedron);
- Remove redundant ions with the largest value of the norm until the number of particles does not equal the required number $N = 12C^3$ (for each cation remove two anions in order to maintain electroneutrality).

As a result, volume ($V$) of the ONC formed by this algorithm will be with high accuracy equal to the volume of CNC with edges of length $L$: $V = L^3 = (aC)^3$. Further, we will use the values of $L$ as the characteristic linear size of CNC and ONC bearing in mind that they are equal in the number of particles and volume.

To plot the size dependences, we conducted a simulation of CNCs at a temperature of 2200K, and ONCs at 2200K and 2300K. This choice of temperatures is due to the fact that the smallest CNC (of 768 ions) melts at a temperature of 2300K, and the corresponding ONC doesn't; on the other hand, relaxation time increases with decreasing temperature. At 2200K the smallest CNC also melts in the beginning of the simulation, but soon re-crystallizes already in octahedral shape.

Table 2 shows the characteristics of CNCs before and after the structural relaxation, relaxation time (before reaching the equilibrium values of the lattice constant and the energy) and total time of MD-simulation. Table 3 shows the characteristics of ONCs, measured after relaxation at two temperatures. Besides, we added to each of these tables the row with extrapolations to crystals of infinite size and the row with results of periodic crystals (768 particles under PBC) simulation from our work [6].

## 4. Size dependence of the nanocrystal lattice parameter

The dependences $a(1/L)$ of the lattice constant upon the reciprocal size of NC after structural relaxation are shown in Fig. 5. It is seen that plots 1 and 2 for ONC fit a straight line in the entire range of sizes. The relaxation time for all the sizes considered does not exceed 1 ns (200 thousand of MD-steps). On the contrary, plot 3 for CNC consists of the two segments. Table 2 shows that the relaxation time increases with the size of CNC, reaching 1 microsecond (200 million of MD-steps), and only the five smallest crystals of the plot 3 have reached an equilibrium. The second segment of plot 3 corresponds to the three largest CNCs, the shape of which did not have time to change during the simulation period.

Let us begin the discussion with the linear dependences obtained for ONCs. At first, we are going to show that decrease of the lattice parameter (with decrease in NC size) is due to compression of the crystals by surface tension.

In the presence of external pressure the following relation holds:
$$a = a_0 (1 - \Delta P / 3K), \qquad (1)$$
where $a$ is the lattice constant of compressed crystal; $a_0$ is the lattice constant at zero external pressure and in the absence of surface tension (e.g. in case of simulation under PBC with a barostat); $K$ is the bulk modulus (the reciprocal of isothermal compressibility).

Then, having the value of $K$ we can estimate the surface overpressure from the formula:
$$\Delta P = 3K (1 - a / a_0) \qquad (2)$$
At the temperature of 2200K the bulk modulus value for the SPP MOX-07 is $K \approx 80$ GPa = 0.5 eV/Å$^3$. Thus, compression of the smallest NC of 768 ions is equivalent to overpressure $\Delta P = 1.5$ GPa = 15000 atm.

Moreover, formula (1) corresponds to the linear dependence of the lattice constant of the reciprocal size. Indeed, there is formula known for spherical surfaces:

$$\Delta P_{\text{sphere}} = 2\gamma / R, \qquad (3)$$
where $\gamma$ is surface tension constant, $R$ is radius of the sphere. Substitution of expression (3) into (1) gives that
$$a = a_0 (1 - (2\gamma / R) / 3K). \qquad (4)$$

It should be noted, that in macroscopic systems surface tension constant $\gamma$ coincides with the surface energy density $\sigma$, since the macroscopic change of surface area is caused by the transition of molecules from the bulk to the surface and back. Moreover, $\sigma$ characterizes the difference between the binding energies of the molecule on the surface and in the bulk. However, in the case discussed above NCs are compressed by the elastic surface energy $\gamma$, which prevents expansion of internal volume up to lattice constant of $a_0$ without changing the number of molecules on the surface. That's why we expect a difference of values of $\gamma$ and $\sigma$ to be obtained in the processing of our data.

Since our model crystals are not spherical, then we can not directly use the formula (4) for the processing of $a(1/L)$ dependences. We need to get a more general formula for the relation of $\Delta P$ and $\gamma$. In this paper we confine ourselves to the isotropic approximation (in which the pressure is scalar), since we have not studied the direction dependence of the nanocrystals lattice constant.

Thus, in equilibrium the change in energy with decreasing surface area must be equal to surface work for compression of crystal with overpressure $\Delta P$:
$$\gamma dS = \Delta P dV, \qquad (5)$$
hence
$$\Delta P = \gamma dS / dV \qquad (6)$$

In particular, for a sphere $V=4\pi R^3/3$, $dV=4\pi R^2 dR$, $S=4\pi R^2$, $dS=8\pi R dR$, where $dS/dV=2/R$, and formula (6) reduces to equation (3). Now, we write expression (3) in terms of L, to compare it with formulas (6) for cubic and octahedral shapes:
$$R = (3/4\pi)^{1/3} L$$
$$\Delta P_{sphere} = \frac{2\gamma}{R} = \frac{2(4\pi/3)^{1/3}}{L}\gamma \approx \frac{3.223984}{L}\gamma \qquad (7)$$

In the case of cubic crystals $V=L^3$, $dV=3L^2$, $S=6L^2$, $dS=12L$, hence
$$\Delta P_{\text{cube}} = 4\gamma / L \qquad (8)$$

In the case of octahedral crystals, dependence of the surface area and volume upon the octahedron edge length $O$ can be derived from the following: $V=(\sqrt{2}/3)O^3$, $S=(2\sqrt{3})O^2$, $dV=(\sqrt{2})O^2 dO$, $dS=(4\sqrt{3})OdO$, hence $dS/dV=2\sqrt{6}/O$. Octahedron edge length is connected with edge length of cube of the same volume $V=(\sqrt{2}/3)O^3=L^3$ by the following relation
$$O = (3^{1/3} / 2^{1/6}) L. \qquad (9)$$
Hence,
$$\Delta P_{octa} = \frac{2\sqrt{6}}{O}\gamma = \frac{2^{5/3}3^{1/6}}{L}\gamma \approx \frac{3.812737}{L}\gamma \qquad (10)$$

It is seen that the factor value of 3.813 in front of the surface tension constant γ in the formula for octahedra lies between the values of 4 and 3.224 for cubes and spheres, closer to the value for cubes. On the other hand, γ values themselves must differ for different crystallographic planes.

After substituting expression (10) into formula (1), it turns out that

$$a = a_0 (1 - 3.812737\, (\gamma / L) / 3K) = \\ = a_0 (1 - 2.541825\, (\gamma / L)) \qquad (11)$$

The results of applying formula (11) to linear plots 1 and 2 for octahedra in Fig. 5 are given in the third and the fourth columns of Table 4. The values of $a_0$ are extrapolations of the lattice constant of nanocrystals to the region of infinite (macroscopic) size and within the tolerance equal to the values obtained under PBC for quasi-infinite periodic crystals. This equality validates the model used.

Now return to the plot 3 in Fig. 5, which corresponds to $T = 2200$K and the crystals having a cubic shape at the beginning of simulation. This plot consists of two sections. Herewith the section for the three largest CNCs (with unfinished structural relaxation) is located above the plot for ONC and has a slightly greater slope. The parameters of its approximation by formula (11) are the following: $a_0 = 5.612 \pm 0.002$ Å and $\gamma = 0.0567 \pm 0.008$ эВ/Å². At the same time, the greater value of γ means that the cubic surfaces {100} are energetically less favorable than the octahedral surfaces {111}. However, due to the small number of points the error of this approximation is high. And the five smaller CNCs due to long enough simulation time had reached equilibrium in which the shape and the lattice parameter coincide with ones for ONC.

Finally, since the lattice constant of CNC after 0.1 ns simulation (i.e. after the kinetic relaxation, but before structural relaxation) did not have time to change from the originally specified macroscopic value of 5.603Å, then the corresponding straight line (which is excluded from Fig. 5) would be a constant (parallel to abscissa) with zero value of γ.

## 5. Size dependence of the nanocrystal specific energy

In order to compare the characteristics of the crystallographic planes {100} and {111}, we measured the energy of NCs before their transformation (i.e., after kinetic relaxation) and after reaching the equilibrium state (i.e., after the structural relaxation). The corresponding reciprocal size dependences of specific energy are drawn in Fig. 6. In this case, the energy values are normalized to macroscopic (without the surface compression) volume $V_0 = L_0^3 = (a_0 C)^3$, where C is the number of unit FCC cells on the edge of CNC, $a_0$ is the macroscopic lattice constant. Thus, the volume $V_0$ is proportional to the number of molecules, allowing the analysis of size dependence of the specific energy of NC separately from the size dependence of the lattice constant.

As it seen from Fig. 6, the reciprocal size (1/$L$) dependences of the specific energy $E/V_0$ for ONCs are again close to straight lines. This behavior corresponds to the influence of the surface energy, the positive contribution of which reduces the absolute value of the total binding energy. In order to isolate the contribution of surface energy one can represent specific internal energy of the crystal of volume $V_0$ and area $S_0$ in the form

$$E = uV_0 + \sigma S_0 \rightarrow E / V_0 = u + \sigma (S_0 / V_0), \qquad (12)$$

where $u$ and σ are the volume and surface energy densities, respectively. Ratio $S_0/V_0$ is proportional to the reciprocal linear size 1/$L$, and if $u$ and σ do not depend on the size, then formula (12) defines a straight line in coordinates $E/V_0 = f(1/L)$.

Let us note again, that in the formula (12) we use the values of area and volume $S_0$ and $V_0$ corresponding to macroscopic period $a_0$, not including the surface compression. The difference between use of volume $V_0$ instead of volume $V$, which takes the compression into account, is that $V_0$ is strictly proportional to the amount of substance (number of molecules, mass of crystal), and $V$ is not. Indeed, from formulas (9)–(10) it follows that taking compression into account leads to equations

$$V = V_0(1 - Const \cdot L^{-1}) \Rightarrow \\ \frac{E}{V} = \frac{E}{V_0(1 - Const \cdot L^{-1})} \approx \frac{E}{V_0}(1 + Const \cdot L^{-1}). \qquad (13)$$

It is seen that division by volume V, which takes account of the compression, would bring into the size dependence of the specific energy an additional term (linear by 1/$L$), which is determined only by the change in volume $V$, but not directly related to energies of molecules in the bulk or on the surface. In our opinion, the experimental data on surface energy of $UO_2$ are determined by energy of molecules on the surface, rather than the ratio of internal energy to the actual volume $V$, so that in order to calculate σ we used expression (12) without correction (13).

For cubes $V_0 = L^3$, $S_0 = 6L^2$, hence $S_0/V_0 = 6/L$, so that formula (12) is transformed into

$$E / V_0 = u + 6\, (\sigma / L) \qquad (14)$$

As shown above, for octahedra with edge length $O$ $V = (\sqrt{2}/3)O^3$ and $S = (2\sqrt{3})O^2$, hence $S/V = (3\sqrt{6})/O$. Substituting (9) here, we obtain $S/V = (3^{7/6} 2^{2/3})/L$. Then expression (12) for octahedra takes the form:

$$E / V_0 = u + 5.719106\, (\sigma / L). \qquad (15)$$

In deriving (15) we, for simplicity, have neglected the fact that corners of our octahedra are smoothed, as the area of these regions is small compared with the total area of the crystal.

The results of applying formulas (14) and (15) to CNCs and ONCs (plots 1, 3 and 4 in Fig. 6) are given in the first and the second columns of Table 4. If the specific energy is normalized to volume $V$, which includes the compression, its size dependence virtually disappears (the slope within the tolerance is zero) and determination of $\sigma$ becomes impossible. However, we do not exclude the possibility that in some experimental techniques of measuring the surface energy its value includes contribution from compression of the bulk volume.

Note also, that for small sizes of NC values $u$ and $\sigma$ could strongly depend on the size, since volume of the surface layer is comparable to the total volume. In our case it turned out that the values of $E/V$ up to the smallest NCs (of 768 ions) lie on the same straight lines as the rest, with no significant deviations. Thus, the values of $u$ and $\sigma$ are close to constant, indicating a relatively small thickness of the surface layer (of the order of one lattice constant).

On the other hand, at both temperatures 2200 K and 2300 K the coefficients of the straight lines fitted to the three smallest ONCs still differ from the coefficients of lines fitted to all points, even taking tolerances into account (three points are the minimum which is necessary to estimate the tolerance). So we give them separately in Table 4. This difference may indicate a nonlinearity of size dependences for small NCs.

The comparison shows that the values of $u$ obtained for CNCs and ONCs of all sizes within the tolerance coincide with the values of specific energy received by us for quasi-infinite periodic crystals under PBC. This demonstrates the correctness of the extrapolation and correspondence of the calculation results to the physical meaning of formulas (12), (14) and (15). Decrease in absolute value of $u$ with temperature increasing from 2200K to 2300K is connected with increasing energy of thermal vibrations, as well as the thermal expansion.

In the modern review IAEA-06 [1] the experimental data for surface energy of $UO_2$ is given from estimates of Hall et al. [17], who has proposed the following recommendations as a result of non-trivial data processing (e.g., use of a reduction factor 1.5 to account for pre-heating of the samples): temperature dependence $\sigma(T)$ lies in a region bounded by the lower value of 0.2 J/m$^2$ and from above by the line

$$\sigma_{max} = 0.0936 - 0.0000176 \, (T - 273) \text{ eV/Å}^2, \quad (16)$$

so that the mean estimate lies on the line

$$\sigma_{mean} = 0.0531 - 0.0000874 \, (T - 273) \text{ eV/Å}^2 \quad (17)$$

Table 5 shows that our energy density for the surface {111} $\sigma_{111} = 0.071$ eV/Å$^2 \approx 1.137$ J/m$^2$ (1 eV/Å$^2 \approx 16.022$ J/m$^2$) within the error corresponds to the experimental data at T = 1770K from Nikolopoulos [18], but it is by ~20% higher compared to $\sigma_{max}$(2200K), and by 2 times higher than $\sigma_{mean}$(2200K). On the other hand, uncertainty of the recommendation is ~70%, which is 3.5 times the difference between $\sigma_{max}$ and our result. And the review [1] questioned the fact of decreasing surface energy with temperature. In addition, there are recent experimental data of Matzke et al. [19] at room temperature, which, are 20% above $\sigma_{max}$, similar to our calculations.

Results of static SPP-calculations of Skomurski et al. [2] with an average value of ~1.1 J/m$^2$ are close to ours, and the average value of Tan et al. [7] ~1.4 J/m$^2$ is higher than our value by 20%. Both results [2] and [7] correspond to zero temperature, and they satisfy the recommended value $\sigma_{mean}$(273K) due to its large uncertainty. However, these results are 60% and 30% below the data of Matzke for $T = 300$K. DFT-calculations of Skomurski are less plausible: the surface energy decreases rapidly with increasing thickness of slab and reaches a value of 0.27±0.13 J/m$^2$, which is 3 times lower than the recommendation and 6 times lower compared with the data of Matzke. In recent accurate DFT-calculations of Evarestov et al. [3] the obtained values are near 0.94 J/m$^2$, which is similar to our results and to the recommendations. The authors [3] have also explained the too low values obtained by Skomurski by pointing out the ferromagnetic state of $UO_2$ modeled in [2] instead of the correct antiferromagnetic state.

We also evaluated the energy density for the surface {100} $\sigma_{100} = 0.1$ eV/Å$^2 \approx 1.602$ J/m$^2$ from the data of Fig. 6, which shows that plot 1 for unrelaxed CNCs lies above the plot for ONCs at the same temperature and has a bigger slope. Thus, the lower surface energy is the main reason for the change of the shape of NCs from cubic to octahedral in the course of prolonged MD simulation.

Values of $\gamma$ (the last column of Table 4) calculated in the previous section can also be compared with surface energy density, since they are in the same units. It is seen that our value $\gamma$(2200K) = 0.0546 eV/Å$^2 \approx 0.875$ J/m$^2$ is closer to the experimental estimates than our value of $\sigma_{111}$ (see Table 5). As we discussed above, there is no reason for equality of $\gamma$ and $\sigma$, since $\gamma$ is a constant of surface mechanical tension, which alters the energy of NC only by changing the distances between ions and not due to changes in the number of molecules on the surface.

## *6. Comparison of surface energy of octahedral and cubic nanocrystals*

It is known that the equilibrium shape of single crystal, which provides a minimum surface energy, is determined by the ratio of surface energies on the crystallographic planes of its constituent faces. Castell et al. [16] experimentally investigated the microscopic bubbles in single crystals of uranium dioxide. The shape of cavities (bubbles) in $UO_2$ monocrystals should be equal to the equilibrium shape of nanocrystals, as it is determined by the same condition of minimum surface energy.

In the experiments of Castell [16] the smallest bubbles (with size of 200–500 nm) had the shape of octahedron formed by planes {111} with corners smoothed by planes {100} (see Fig. 4). The author measured the ratio of areas of planes {111} and {100}: $\omega = S_{100}/S_{111} = 0.0616\pm0.0194$, and then he determined the ratio of surface energy densities for these planes as

$$\frac{\sigma_{100}}{\sigma_{111}} = \sqrt{3} - \sqrt{\frac{\sqrt{3}\omega}{1+\sqrt{3}\omega}} = 1.42 \pm 0.05$$

In our simulations NCs also take the shape of octahedron with smooth corners after structural relaxation (see Fig. 3), and the corresponding ratio $\sigma_{100}/\sigma_{111} = 1.408 \pm 0.042$ within a tolerance coincides with the experimental result of Castell.

In comparison with our modeling of non-periodic NCs, surfaces of which are formed naturally, all the previous calculations used 2D [2] [3] [7] or 3D [2] periodicity, which leads to a bunch of artificial problems.

For example, in the case of 2D-periodicity the Ewald sum converges slower, and the relaxation by lattice statics method requires fixing positions of some atoms, which represent the bulk region of the crystal [7]. However, 3D-periodicity leads to an interaction between the two surfaces of the slab (due to translated reflections), and the reciprocal thickness dependence of surface energy for DFT-calculations of Skomurski et al. [2] doesn't saturate, but decreases to minus infinity (as opposed to our reciprocal size dependences for NCs).

As a result, ratios $\sigma_{100}/\sigma_{111}$ in all the previous calculations were greatly overestimated, especially in the case of 3D-periodicity (see Table 5).

Many authors also note that in simulations under PBC the surface {100} is unstable due to the dipole moment [2]. Tan et al. [7] tried to minimize its energy searching through a large number of possible configurations, as well as modeling it with trenches in order to reduce the ratio $\sigma_{100}/\sigma_{111}$ (in accordance with the proposal of Castell), but the gain was only 12–14%. In the most recent and accurate DFT-calculations of Evarestov et al. [3] the surface {100} hasn't been considered at all.

In addition, SPP-calculations in the works of Tan [7] and Skomurski [2] were carried out with outdated pair potentials for the shell model, which poorly reproduce the thermophysical experimental data for $UO_2$ (see reviews [4] [5]). And this also could be the cause of big discrepancy between their results and Castell's.

### 7. Bulk and surface energy before and after kinetic relaxation

In order to assess the impact of relaxation, we have drawn reciprocal size dependences of the specific energy of ONCs and CNCs at the first step of MD-simulation (prior to any relaxation). At this time we used a simpler algorithm for ONC construction of unit FCC cells, which produces a step-wise surface with area greater than area of flat surface resulting from the algorithm described in section 3. In Fig. 7 the data for non-relaxed octahedra lie above the corresponding data for cubes because of the excess energy of such step-wise surface. However, even after a brief kinetic relaxation (t = 0.1 ns), the plot for ONC lie below the plot for CNC.

Applying formulas (14) and (17) to the data for unrelaxed crystals we got the values $u = -1.08376\pm0.00004$ eV/Å$^3$, $\sigma = 0.3520\pm0.0008$ eV/Å$^2$ for ONCs and $u = -1.08444\pm0.00002$ eV/Å$^3$, $\sigma = 0.2241\pm0.0004$ eV/Å$^2$ for CNCs. As expected, the values of $u$ coincide with the extrapolations in Table 4 up to the fourth digit, since the relative contribution of relaxation tends to zero with increasing size. But the values of the surface energy density $\sigma$ without relaxation were overstated by 2–5 times, which demonstrates the importance of kinetic relaxation.

### 8. Specific heat capacity

Calculation of the crystal energies at two close temperatures 2200 K and 2300 K allowed us to determine the values of isobaric heat capacity of model nanocrystals at the mean temperature of 2250 K using the relation $C_p = \Delta E/\Delta T$.

Size dependence of the specific energy is determined by change of the ratio of bulk and surface ions. As in one phase the shape of nanocrystal does not depend on temperature and the surface area alters insignificantly, this ratio is conserved. Consequently, the finite difference of the crystal energy at close temperatures is almost independent of size (see Table 3 and Fig. 6).

Extrapolation of the heat capacity to crystals of infinite size in this work is found to be 114±1 J/(mol K), which is a little higher than a value of 108 J/(mol K), obtained in our simulation under PBC [6]. However, in that work we have simulated quasi-infinite periodic crystals without a surface, so in the comparison with experimental specific heat we compensated the lack of Schottky defects (which are formed on the surface) by empirical formula of Hyland and Ralph [20]. Taking into account this contribution to the heat capacity measured under PBC at 2250 K, we got the value of 115 J/(mol K), which is within tolerance identical to the result of this work.

### 9. Conclusions

Compared with the previous works on computer simulation of surface of uranium dioxide ($UO_2$), this work features the first time use of the graphics processing units (GPU) with NVIDIA CUDA technology and the method of molecular dynamics (MD) instead of the lattice statics; nonperiodic nanocrystals (NCs) of volume up to 1000 nm$^3$ (50000 particles) instead of periodic crystals (of 100–1000 particles); the empirical pair potentials MOX-07 [8] in the approximation of rigid ions (which reproduce well the large specter of experimental data for $UO_2$ [6]) instead of the shell model or quantum-mechanical (ab initio) calculations by density functional theory (DFT).

It is shown that the cubic shape of NC is metastable, and equilibrium is reached in the process of structural relaxation to the octahedral shape during the time of 100–1000 ns (200 million MD steps), which increases with the size of NC. This result coincides with the results of natural experiments of Castell [16], where microscopic cavities (bubbles) in single crystals of $UO_2$ also took the form of octahedra (with smooth corners).

We obtained and analyzed the reciprocal size dependences of the lattice constant and specific energy for NCs of cubic and octahedral shape at two temperatures 2200 K and 2300 K (because our smallest cubic NC melts at a temperature of 2300 K, and the corresponding octahedral NC does not; on the other hand, the relaxation time increases with decreasing temperature). It is shown that plots of these dependences are close to linear, and their extrapolation to the region of macroscopic (infinite) crystals give values close to those estimated for quasi-infinite periodic crystals. In addition, we showed that the volume and surface energy density and the heat capacity (calculated by the difference of energies at temperatures of 2200 K and 2300 K) are practically independent of NC size, indicating that the surface layer has small thickness (of the order of one lattice constant).

At a temperature of 2200 K, our value of surface energy density for octahedral surface {111} of 1.137±0.032 J/m$^2$ exceeds the upper bound of 0.96 J/m$^2$ from Hall's review [17] by 20% and the recommended value of 0.58 J/m$^2$ by 2 times, and our surface tension constant of 0.875±0.008 J/m$^2$ lies between them. But Hall's recommendations are based on 30 years old experiments and their uncertainty of 70% significantly exceeds the discrepancy with our result. In addition, the experimental data of Matzke [19] are also 20% higher than the Hall's upper bound, and Evarestov et al. in the recent DFT-calculations [3] for octahedral surface {111} got a value of 0.94 J/m$^2$, which is also close to our results.

The ratio of energy density on surface {100} of cubes and on surface {111} of octahedra was found to be 1.408±0.042, which within the tolerance coincides with a value of 1.42±0.05 from the experimental work of Castell [16]. This result corresponds to the experiment better than all the previous calculations, which yielded values in the range of 1.7–3.9 [2] [7].

## References

1. Thermophysical Properties Database of Materials for Light Water Reactors and Heavy Water Reactors. IAEA (2006). http://www-pub.iaea.org/MTCD/publications/PDF/te_1496_web.pdf
2. F.N. Skomurski, R.C. Ewing, A.L. Rohl, J.D. Gale, U. Becker. American mineralogist 91 (2006) 1761–1772.
3. R. Evarestov, A. Bandura, E. Blokhin. Acta Materialia 57 (2009) 600–606.
4. K. Govers, S. Lemehov, M. Hou, M. Verwerft. Journal of Nuclear Materials 366 (2007) 161–177.
5. K. Govers, S. Lemehov, M. Hou, M. Verwerft. Journal of Nuclear Materials 376 (2008) 66–77.
6. S.I. Potashnikov, A.S. Boyarchenkov, K.A. Nekrasov, A.Ya. Kupryazhkin. http://arxiv.org/abs/1102.1529
7. A.H.H. Tan, M. Abramowski, R.W. Grimes, S. Owens. Physical Review B 72 (2005) 035457.
8. S.I. Potashnikov, A.S. Boyarchenkov, K.A. Nekrasov, A.Ya. Kupryazhkin, ISJAEE 8 (2007) 43. http://isjaee.hydrogen.ru/pdf/AEE0807/AEE08-07_Potashnikov.pdf
9. http://en.wikipedia.org/wiki/Buckingham_potential
10. http://en.wikipedia.org/wiki/Semi-implicit_Euler
11. H. Berendsen, J. Postma, W. Van Gunsteren, A. Haak, J. Dinola. Journal of Chemical Physics 81 (1984) 3684–3690.
12. A.S. Boyarchenkov, S.I. Potashnikov, Numerical methods and programming 10 (2009) 9. http://num-meth.srcc.msu.ru/english/zhurnal/tom_2009/v10r102.html
13. A.S. Boyarchenkov, S.I. Potashnikov, Numerical methods and programming 10 (2009) 158. http://num-meth.srcc.msu.ru/english/zhurnal/tom_2009/v10r119.html
14. B.M. Shaub. American Mineralogist 23 (1938) 334–341. http://www.minsocam.org/ammin/AM23/AM23_334.pdf
15. A. Portnoff-Porneuf. Journal of Nuclear Materials 2 (1960) 186–188.
16. M.R. Castell. Physical Review B 68 (2003) 235411.
17. R.O.A. Hall, M.J. Mortimer, D.A. Mortimer. Journal of Nuclear Materials 148 (1987) 237–256.
18. O. Nikolopoulos, B. Schulz. Journal of Nuclear Materials 82 (1979) 172–178.
19. H. Matzke, T. Inoue, R. Warren. Journal of Nuclear Materials 91 (1980) 205–220.
20. G.J. Hyland, J. Ralph. High Temperatures High Pressures 15 (1983) 179–190.

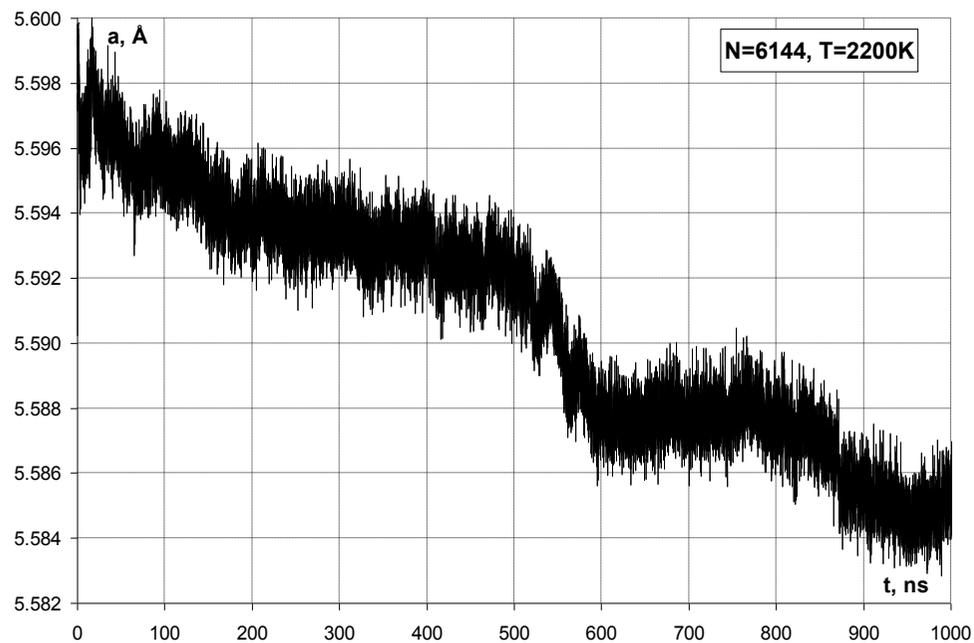

FIG. 1. Decrease of lattice parameter during prolonged simulation (1 microsecond) due to transformation of NC from the cubic shape to octahedral.

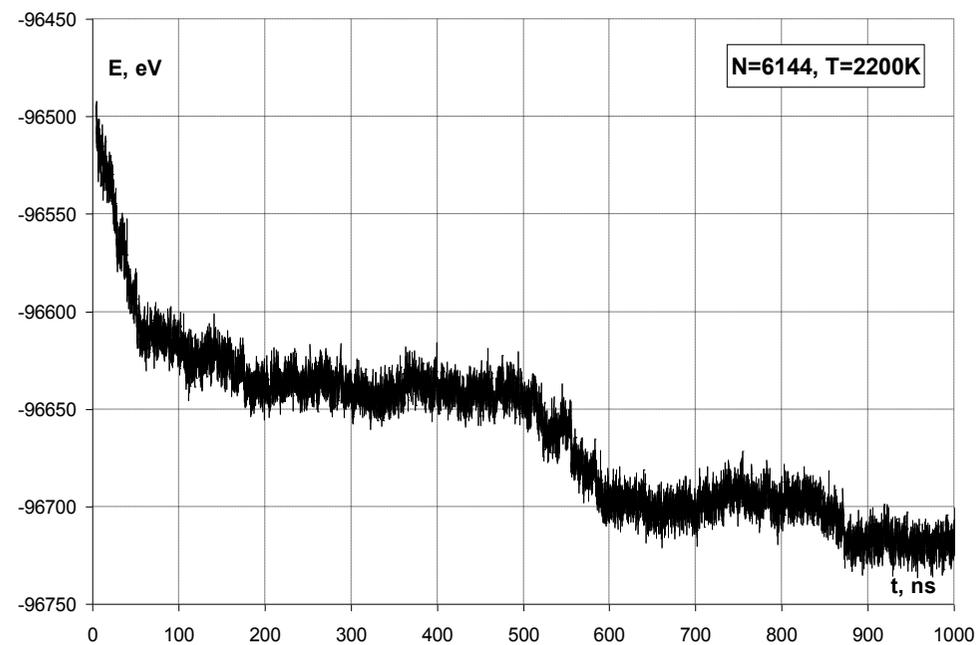

FIG. 2. Decrease of energy during prolonged simulation (1 microsecond).

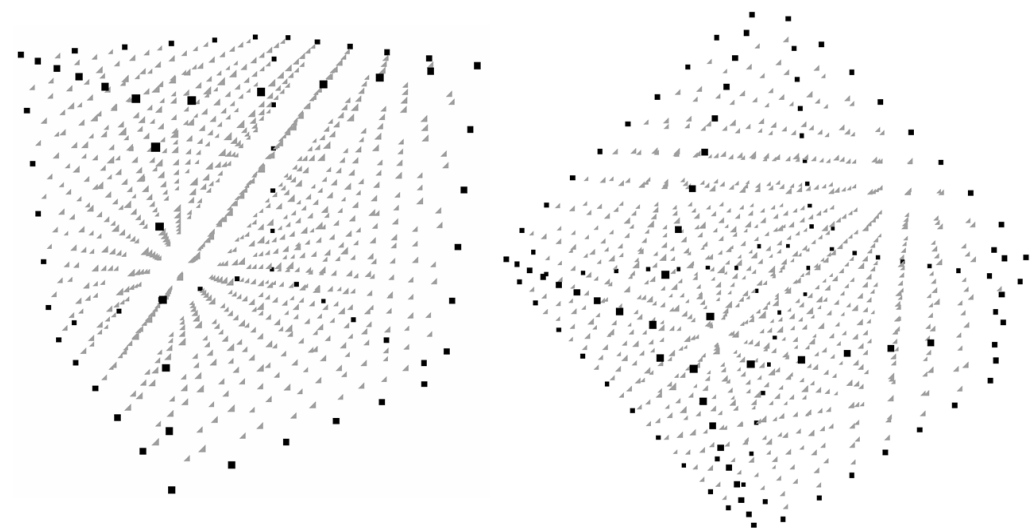

FIG. 3. Uranium sublattice of NC of 2592 ions before and after 1 microsecond structural relaxation.

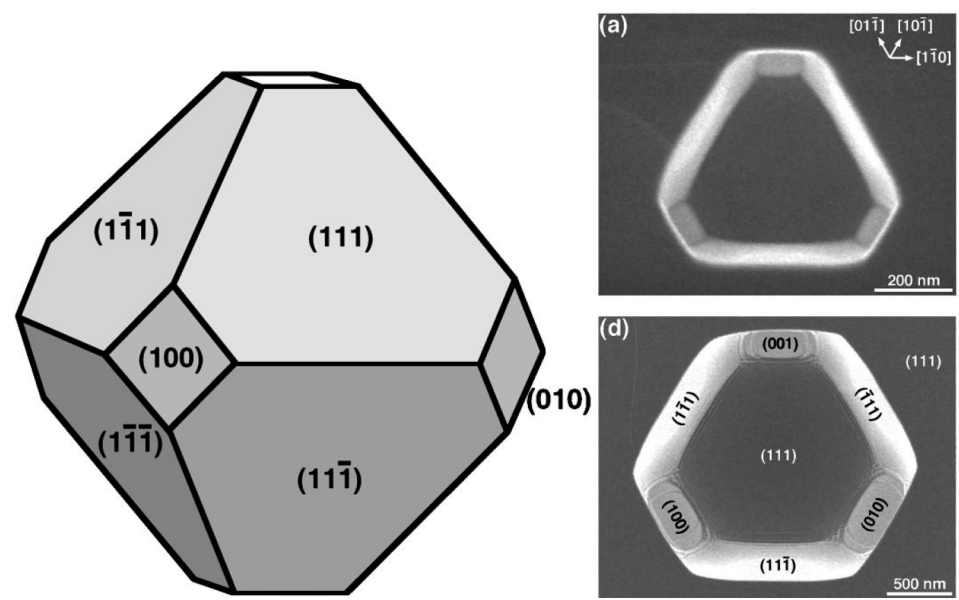

FIG. 4. Shape of microscopic cavities in $UO_2$ from Castell's work [16].

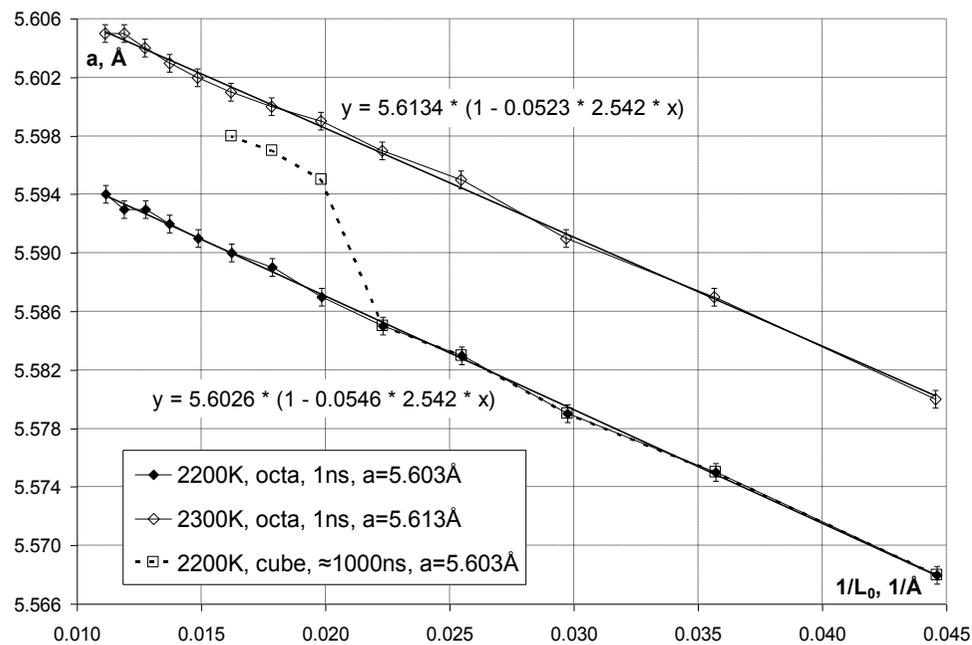

FIG. 5. NC reciprocal size dependence of lattice parameter.

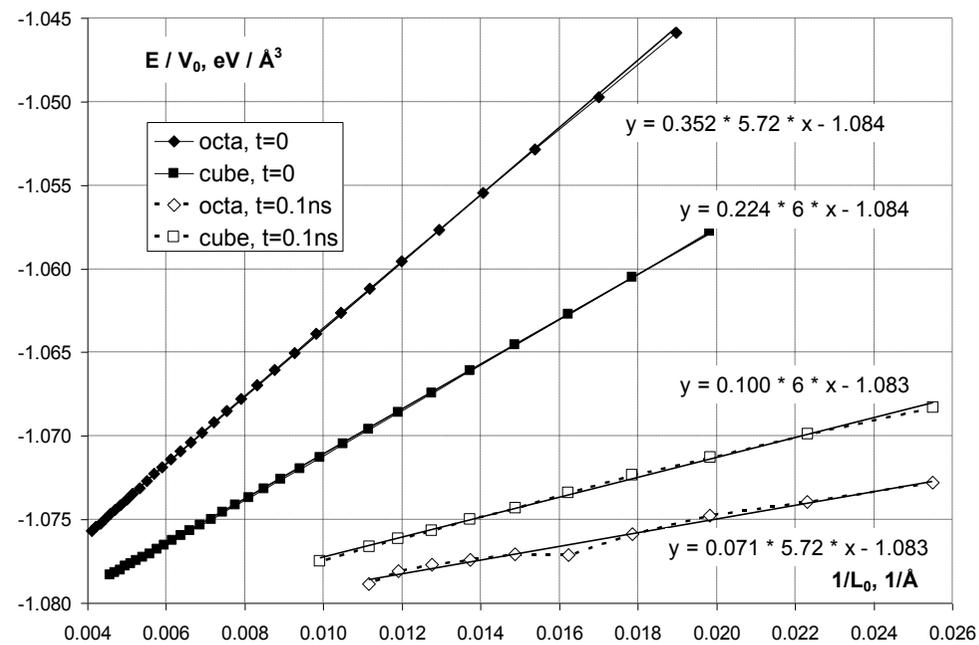

FIG. 7. Specific energy of cubic and octahedral NCs before and after ~0.1 ns kinetic relaxation.

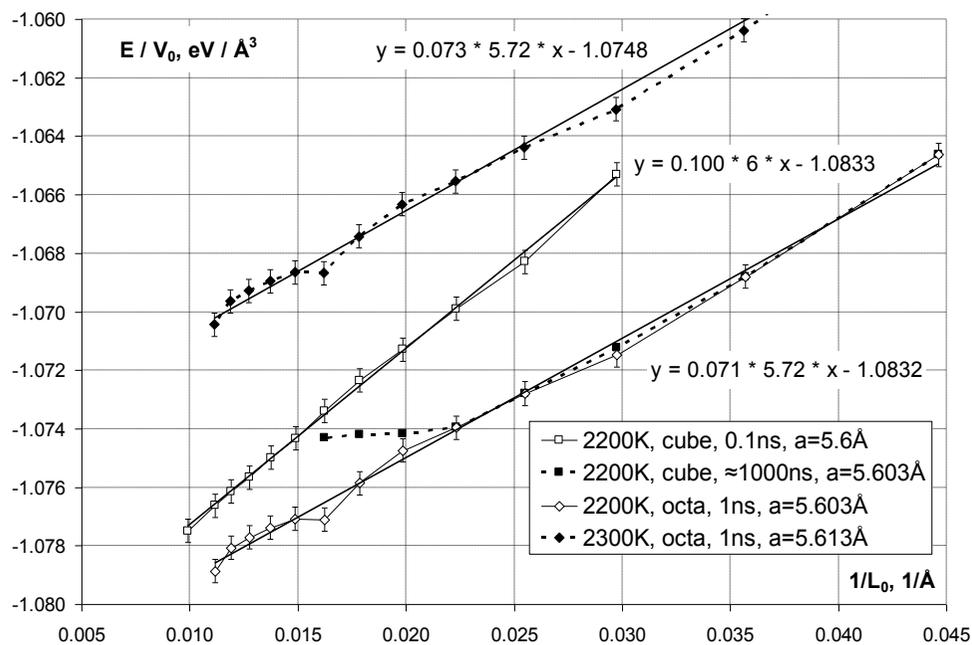

FIG. 6. NC reciprocal size dependence of specific energy.

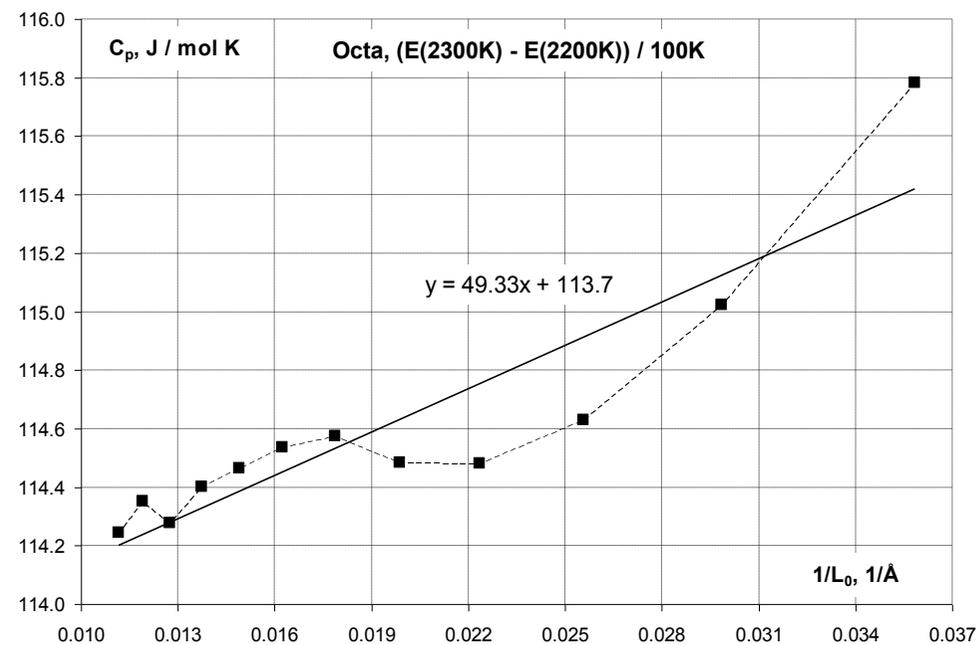

FIG. 8. NC reciprocal size dependence of specific heat capacity at 2250 K.

TABLE 1. Ionicity $Q$ and short-range interaction parameters for our SPP MOX-07.

| Compound | $Q$ | $X$--, eV | $Y$--, 1/Å | $Z$--, eV*Å$^6$ | $X$+-, eV | $Y$+-, 1/Å |
|---|---|---|---|---|---|---|
| $UO_2$ | 0.68623 | 50211.74 | 5.52 | 74.7961 | 873.107 | 2.783855 |
| $PuO_2$ | 0.68623 | 50211.74 | 5.52 | 74.7961 | 871.790 | 2.807875 |

TABLE 2. Characteristics of the cubic-shaped nanocrystals at T = 2200K before and after the structural relaxation

| Ion count | $L_0$, Å | * $E_{0.1}/V_0$, eV/Å$^3$ | $E/V_0$, eV/Å$^3$ | $a$, Å | $P$, GPa | Structural relaxation time, ns | Total MD time, ns |
|---|---|---|---|---|---|---|---|
| 768 | 22.41 | ** −1.0570 | −1.0646 | 5.568 | 1.499 | 5 | 320 |
| 1500 | 28.02 | −1.0628 | −1.0688 | 5.575 | 1.199 | 600 | 900 |
| 2592 | 33.62 | −1.0653 | −1.0712 | 5.579 | 1.028 | 500 | 1000 |
| 4116 | 39.22 | −1.0683 | −1.0728 | 5.583 | 0.857 | 220 | 700 |
| 6144 | 44.82 | −1.0699 | −1.0740 | 5.585 | 0.771 | 900 | 1070 |
| 8748 | 50.43 | −1.0713 | −1.0742 | 5.595 | 0.343 | – | 360 |
| 12000 | 56.03 | −1.0723 | −1.0742 | 5.597 | 0.257 | – | 32 |
| 15972 | 61.63 | −1.0734 | −1.0743 | 5.598 | 0.214 | – | 21 |
| Infinity | – | −1.0833 | −1.0835 | 5.6022 | – | – | – |
| 768 PBC | 22.41 | −1.0830 | – | 5.6028 | – | – | 0.1 |

* $E_{0.1}$ – energy after 0.1 ns of simulation, i.e. before the structural relaxation but after the kinetic relaxation; ** – melt.

TABLE 3. Characteristics of the octahedral nanocrystals after the structural relaxation

| Ion count | $L_0$, Å | | $E/V_0$, eV/Å$^3$ | | $a$, Å | | $P$, GPa | | $C_p$(2250K), kJ/mol K |
|---|---|---|---|---|---|---|---|---|---|
| | 2200K | 2300K | 2200K | 2300K | 2200K | 2300K | 2200K | 2300K | |
| 768 | 22.41 | 22.45 | −1.0646 | −1.0559 | 5.568 | 5.58 | 1.499 | 1.411 | 0.13191 |
| 1500 | 28.02 | 28.07 | −1.0688 | −1.0604 | 5.575 | 5.587 | 1.199 | 1.112 | 0.11578 |
| 2592 | 33.62 | 33.68 | −1.0715 | −1.0631 | 5.579 | 5.591 | 1.028 | 0.941 | 0.11502 |
| 4116 | 39.22 | 39.29 | −1.0728 | −1.0643 | 5.583 | 5.595 | 0.857 | 0.770 | 0.11814 |
| 6144 | 44.82 | 44.90 | −1.0740 | −1.0655 | 5.585 | 5.597 | 0.771 | 0.684 | 0.11448 |
| 8748 | 50.43 | 50.52 | −1.0747 | −1.0663 | 5.587 | 5.599 | 0.685 | 0.599 | 0.11448 |
| 12000 | 56.03 | 56.13 | −1.0758 | −1.0674 | 5.589 | 5.600 | 0.600 | 0.556 | 0.11458 |
| 15972 | 61.63 | 61.74 | −1.0771 | −1.0687 | 5.590 | 5.601 | 0.557 | 0.513 | 0.11454 |
| 20736 | 67.24 | 67.36 | −1.0771 | −1.0686 | 5.591 | 5.602 | 0.514 | 0.470 | 0.11446 |
| 26364 | 72.84 | 72.97 | −1.0774 | −1.0689 | 5.592 | 5.603 | 0.471 | 0.428 | 0.11440 |
| 32928 | 78.44 | 78.58 | −1.0777 | −1.0693 | 5.593 | 5.604 | 0.428 | 0.385 | 0.11428 |
| 40500 | 84.05 | 84.20 | −1.0781 | −1.0696 | 5.593 | 5.605 | 0.428 | 0.342 | 0.11435 |
| 49152 | 89.65 | 89.81 | −1.0789 | −1.0704 | 5.594 | 5.605 | 0.386 | 0.342 | 0.11425 |
| Infinity | – | – | −1.0832 | −1.0748 | 5.6026 | 5.6134 | – | – | 0.11370 |
| 768 PBC | 22.41 | 22.45 | −1.0830 | −1.0744 | 5.6028 | 5.6134 | – | – | 0.11478 |

TABLE 4. The coefficients of linear extrapolation of the reciprocal nanocrystal size dependences of the specific energy and the lattice constant to infinity.

| $T$, K | $u$, eV/Å$^3$ | $\sigma$, eV/Å$^2$ | $a_0$, Å | * $\gamma$, eV/Å$^2$ |
|---|---|---|---|---|
| **Through CNCs of all sizes after 0.1 ns simulation** | | | | |
| 2200 | −1.0833 ± 0.0001 | 0.100 ± 0.001 | 5.603 | – |
| **Through ONCs of all sizes after relaxation** | | | | |
| 2200 | −1.0832 ± 0.0002 | 0.071 ± 0.002 | 5.6026 ± 0.0002 | 0.0546 ± 0.0005 |
| 2300 | −1.0748 ± 0.0002 | 0.073 ± 0.002 | 5.6134 ± 0.0002 | 0.0523 ± 0.0006 |
| **Through the three smallest ONCs after relaxation** | | | | |
| 2200 | −1.0852 ± 0.0002 | 0.081 ± 0.001 | 5.601 ± 0.001 | 0.052 ± 0.002 |
| 2300 | −1.0776 ± 0.0006 | 0.085 ± 0.003 | 5.613 ± 0.001 | 0.052 ± 0.002 |
| **Values from simulation under PBC** | | | | |
| 2200 | 1.0833 ± 0.0002 | – | 5.6028 ± 0.0002 | – |
| 2300 | 1.0745 ± 0.0002 | – | 5.6134 ± 0.0002 | – |

* – values of $\gamma$ were calculated using bulk modulus $K \approx 80$ GPa = 0.5 eV/Å$^3$ measured for our SPP MOX-07 in the temperature range 2200–2300 K, while the specified tolerance of $\gamma$ is only a lower bound, since we have not calculated the uncertainty of $K$.

TABLE 5. Comparison of {111} and {100} surface energy values and their ratio.

| Source | Type of measurement | $T$, K | $\sigma_{100}$, J/m$^2$ | $\sigma_{111}$, J/m$^2$ | $\sigma_{100}/\sigma_{111}$ |
|---|---|---|---|---|---|
| This work | SPP, MD, IBC | 2200 | 1.602 ± 0.016 | 1.137 ± 0.032 | 1.408 ± 0.042 |
| Tan-05 [7] | SPP, LS, 2D-PBC | 0 | 2.42–3.08 | 1.27–1.54 | 1.91–2.08 |
| * Skomurski-06 [2] | SPP, LS, 2D-PBC | 0 | 1.71–2.58 | 0.93–1.31 | 1.80–2.34 |
|  | SPP, LS, 3D-PBC | 0 | 1.77–2.53 | 0.86–1.19 | 1.95–2.65 |
| Skomurski-06 [2] | DFT (PW GGA), 3D-PBC | 0 | 1.04–1.21 | 0.27–0.59 | 2.05–3.85 |
| Evarestov-09 [3] | DFT (LCAO), 2D-PBC | 0 | – | 0.93–0.94 | – |
| IAEA-06 [1] | Hall's recommendation [17] | 273 | – | 0.85 ± 0.65 | – |
|  |  | 2200 | – | 0.58 ± 0.38 | – |
| Matzke-80 [19] | Experiments | 300 | – | 1.8 ± 0.3 | – |
| Nikolopoulos-79 [18] |  | 1770 | – | 0.9 ± 0.2 | – |
| Castell-03 [16] |  | 300 | – | – | 1.42 ± 0.05 |

* – surface energy in the SPP-calculations of Skomurski [2] depends linearly on the reciprocal thickness of the layer, so we extrapolated their values to infinity, similar to our dependences on the reciprocal nanocrystal size.